      [1999/12/01 v1.4c Il Nuovo Cimento]
\documentclass[varenna]{cimento}

\usepackage{graphicx}
\title{Detection of fast neutrons and digital pulse-shape\\ discrimination between neutrons and $\gamma$ rays}
\author{P.-A. S{\"o}derstr{\"o}m}
\institute{Department of Nuclear and Particle Physics\\
	Uppsala University\\
	751 21 Uppsala, Sweden}
\shortauthor{P.-A. S{\"o}derstr{\"o}m}
\shorttitle{Detection of fast neutrons and digital PSD between neutrons and $\gamma$ rays}
\begin{document}

\maketitle

\begin{abstract}
The basic principles of detection of fast neutrons with liquid scintillator detectors are reviewed, together with a real example in the form of the Neutron Wall array. Two of the challenges in neutron detection, discrimination of neutrons and $\gamma$ rays and identification of cross talk between detectors due to neutron scattering, are briefly discussed, as well as possible solutions to these problems. The possibilities of using digital techniques for pulse-shape discrimination are examined. Results from a digital and analog versions of the zero cross-over algorithm are presented. The digital pulse-shape discrimination is shown to give, at least, as good results as the corresponding analogue version.
\end{abstract}

\section{Introduction}

In experimental studies of the structure of exotic nuclei far from the line of $\beta$ stability it is important to accurately and efficiently be able to identify the nuclei produced in the reactions. One of the most common type of reactions used is the heavy-ion fusion-evaporation reaction, in which a number of neutrons, protons and/or $\alpha$ particles are evaporated, and exotic nuclei are produced with very small cross sections. The protons and $\alpha$ particles are for example detected by highly efficient silicon arrays. For studies of proton-rich nuclei, the clean detection of the number of emitted neutrons in each reaction, is of utmost importance. Due to the uncharged nature of the neutrons, they are however much more difficult to detect with high efficiency, than the light charged particles. The future neutron detection arrays will operate at radioactive ion-beam facilities, like SPIRAL2 \cite{spiral} and FAIR \cite{fair}. The demand on these arrays will be much larger due to the intense $\gamma$-ray background originating from the radioactive beam itself.

\section{Neutron detection\label{sec:detection}}

\begin{figure}
 \centering
 \includegraphics[width=0.7\columnwidth, bb=0 0 567 283]{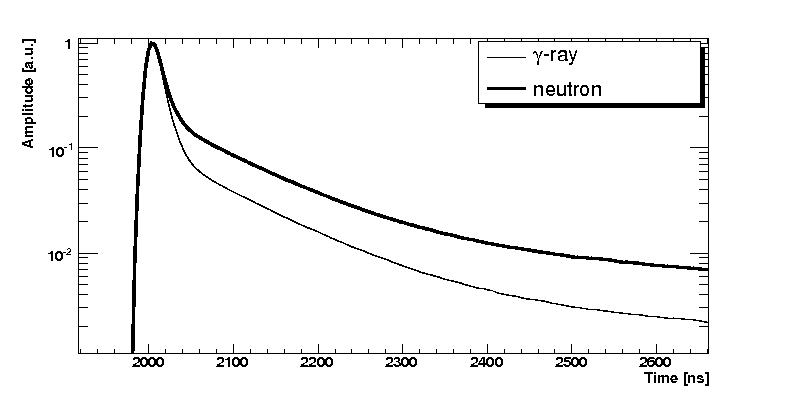}
 \caption{Idealized pulse shapes from a BC501 liquid scintillator from a $\gamma$ ray and a neutron interaction.\label{fig:ngamma}}
\end{figure}

The most common fast neutron\footnote{In this paper, fast neutrons in the energy range 0.2 to 10~MeV are considered.} detector for in in-beam $\gamma$-ray spectroscopy experiments is the organic, liquid scintillator detector. Detection of fast neutrons in an organic scintillator is an indirect process, in which the neutrons deposit energy mainly by elastic scattering with the protons in the liquid. The energy of the recoil protons can have values from zero up to the incoming neutron energy, depending on the scattering angle. The recoil proton excites the organic molecules into either singlet, $S_i$, or triplet states, $T_i$, of order $i$. When the molecule de-excites and cascades down the states it will emit photons, and the decay $S_1 \to S_0$ will be detected as scintillation light in the photomultiplier tube, as a fast component of the pulse. If two molecules in a triplet state collide they can change their configuration to one $S_0$ and one $S_1$ state. The decay of the $S_1$ state contributes with a slow component to the pulse \cite{knoll,birks}. The amount of light produced in this process is proportional to the ionization density of the interacting particle. Hence a proton (neutron) interaction will give a larger contribution of this slow component compared to an electron ($\gamma$ ray). See fig. \ref{fig:ngamma}.

An existing and so far very successful, neutron detector array for use in nuclear structure experiments is the Neutron Wall \cite{1999NIMPA.421..531S}. It consists of 15 hexagonal detectors and one pentagonal, which are assembled into a closely packed array covering about $1\pi$  of the solid angle, and mounted in the forward hemisphere of the setup. The 16 detectors are in turn divided into 50 segments and filled with liquid scintillator of type BC501A \cite{bc501a}. The total volume of the liquid in the Neutron Wall is 150 litre. The Neutron Wall was originally designed for experiments together with EUROBALL \cite{simpseuro}, but since 2005 it is located at GANIL where it is used together with EXOGAM \cite{exogam}. The total neutron efficiency of the Neutron Wall is about 25 \% in symmetrical fusion-evaporation reactions. The pulse-shape discrimination (PSD) electronics used are NIM units of the type NDE202 \cite{NDE202} based on the zero cross-over discrimination method.

\section{Cross talk\label{sec:multi}}

Since the recoil proton only gets a fraction of the full energy of the neutron, the neutron will in many cases have substantial kinetic energy left after the interaction, but a velocity in some random direction. Because of this there exists a probability that the neutron will scatter into a neighboring detector and interact. This can cause quite severe errors in the counting of the number of neutrons emitted in the reaction. Several methods attempting to correct for this exist \cite{1997NIMPA.385..166C,2004NIMPA.528..741}. For the Neutron Wall a method based on the Time-of-Flight difference between the different segments has shown to give good results \cite{2004NIMPA.528..741}. The idea of this rejection technique is that the time difference between the signals of two separate neutrons, on average, is shorter than the time it takes for a scattered neutron to travel between two detectors.

\section{Discrimination between $\gamma$ rays and neutrons\label{sec:psa}}

Another important property of a neutron detector is the ability to discriminate between detection of $\gamma$ rays and neutrons. It has also been shown that even a small amount of $\gamma$ rays mis-interpreted as neutrons dramatically reduce the quality of the cross-talk rejection \cite{2004NIMPA.528..741}. The differences in pulse shape makes it possible to distinguish between neutrons and $\gamma$ rays that hit the detector, and several methods to do this are available. Two well known techniques are the zero cross-over method where the signal is shaped into a bipolar pulse from which the zero crossing is extracted. This shaping can for example be done via a RC-CR integrator and differentiator network \cite{roush1964psd}, or by using double delay lines \cite{alexander1961}. Another technique is the charge comparison method, where the charge in the fast component is integrated and compared to the charge in the slow component \cite{brooks}. Both of these methods are well studied and have been compared using analogue electronics \cite{1995NIMPA.360..584W}.

\section{Digital pulse-shape discrimination\label{sec:newtech}}

When developing the next generation of neutron detector arrays, it will be desirable to use digital instead of analogue electronics. Some of the advantages of using digital electronics is the possibility to store the data and do very complex data treatment off-line, as well as doing advanced digital PSD on-line. For example, instead of using the regular charge comparison method one can enhance the discrimination properties by using a suitable weight \cite{gattimartini} for the integral of the slow component, instead of the very limited modifications to the integration that is available in the analogue case.

\begin{figure}
 \centering
 \includegraphics[width=\columnwidth, bb=0 0 567 198]{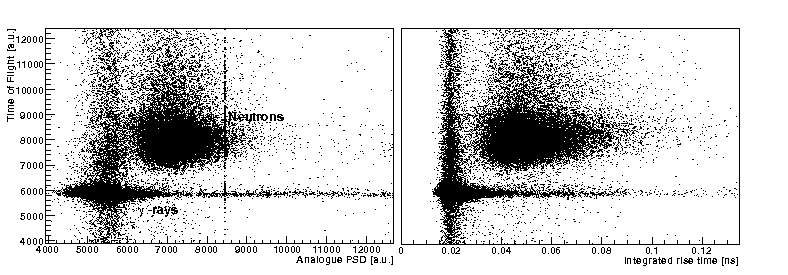}
 \caption{Neutron and $\gamma$-ray separation using analogue (left panel) and digital (right panel) methods in the energy range of 90 to 680 keV for $\gamma$-rays and 710 to 2700 keV for recoil protons. The vertical and horizontal $\gamma$-ray distributions are random coincidences pile-up events, respectively.\label{fig:sep}}
\end{figure}

The digital PSD methods has been tested using an experimental setup consisting of a $^{252}$Cf source, a neutron detector similar to the Neutron Wall detectors and a sampling ADC system running at 300 MHz and 14 bits. As an example, the implementation of a digitized version of the zero cross-over method, the integrated rise-time, is shown in fig.~\ref{fig:sep} together with the analogue separation. In this method of separation, instead of taking the detour with pulse shaping, the information was directly obtained from integrating the pulse and using the time difference between the 10~\% and 72~\% pulse height, which gives information equivalent to the zero cross-over method \cite{nima354_380}. The analogue separation is obtained by one of the Neutron Wall PSD units \cite{NDE202}. By using well known analog algorithms the results can easily be compared with their analogue counterparts, and thus be used for benchmarking of the effectiveness of the digital electronics, especially with respect to the bit resolution and sampling frequencies of the ADC, a work that is currently in progress \cite{results}. As seen in fig.~\ref{fig:sep}, it is possible to get, at least, as good separation with this setup as when using analogue electronics.

\section{Summary and outlook\label{sec:summ}}

A short introduction to the principles of neutron detection by organic liquid scintillation counting has been presented, as well as an example of an existing neutron detection array that has been extensively used. Two of the great challenges in neutron detection, cross talk and  neutron-$\gamma$ discrimination, were also briefly described.

It has been shown that one can get, at least, as good separation of neutrons and $\gamma$ rays using digital and analogue versions of the same type of algorithms. By taking full advantage of the potential of digital PSD it might be possible to get a much better discrimination. For example, nice separation has recently been obtained with a folding approach \cite{2003NIMPA.497..467K} and using a curve fitting method \cite{2002NIMPA.490..299T}. Other possible improvements for the future neutron detector arrays might be the development of new detector materials, like solid plastic scintillators with PSD properties, or deuterated scintillators with better energy response than standard liquid scintillators.

\acknowledgments
Thanks to my supervisor \NAME{J.~Nyberg} for all the help, \NAME{J.~Ljungvall} for the TNT2 library, and to \NAME{R.~Wolters} for the preparation of the setup. I would also like to thank the organizers for a very rewarding summer school. And of course my fellow students for a great time.

\end{document}